\documentclass[a4paper,12pt]{article}
\pdfoutput=1
\usepackage[utf8]{inputenc}
\usepackage[T1]{fontenc}
\usepackage{amssymb}
\usepackage{amsmath}
\usepackage{graphicx}
\usepackage{color}

\DeclareTextSymbol{\degre}{T1}{23}

\begin{document}

\title{\Large \bf Present and Last Glacial Maximum climates as states of maximum entropy production}

\author{\large C. Herbert$^{ab}$\thanks{Email: corentin.herbert@lsce.ipsl.fr}, D. Paillard$^{a}$, M. Kageyama$^{a}$ and B. Dubrulle$^{b}$\\
\\
\small $^a$Laboratoire des Sciences du Climat et de l'Environnement,\\
\small IPSL, CEA-CNRS-UVSQ, Gif-sur-Yvette, France.\\
\small $^b$Service de Physique de l'Etat Condens\'e,\\
\small DSM, CEA Saclay, CNRS URA 2464, Gif-sur-Yvette, France. }
\date{\small January 13, 2011}

\maketitle

\begin{abstract}
\small
The Earth, like other planets with a relatively thick atmosphere, is not locally in radiative equilibrium and the transport of energy by the geophysical fluids (atmosphere and ocean) plays a fundamental role in determining its climate. Using simple energy-balance models, it was suggested a few decades ago that the meridional energy fluxes might follow a thermodynamic \emph{Maximum Entropy Production} (MEP) principle. 
In the present study, we assess the MEP hypothesis in the framework of a minimal climate model based solely on a robust radiative scheme and the MEP principle, with no extra assumptions. Specifically, we show that by choosing an adequate radiative exchange formulation, the \emph{Net Exchange Formulation}, a rigorous derivation of all the physical parameters can be performed. The MEP principle is also extended to surface energy fluxes, in addition to meridional energy fluxes.
The climate model presented here is extremely fast, needs very little empirical data and does not rely on \emph{ad hoc} parameterizations. We investigate its range of validity by comparing its performances for pre-industrial climate and Last Glacial Maximum climate with corresponding simulations with the IPSL coupled atmosphere-ocean General Circulation Model \emph{IPSL\_CM4}, finding reasonable agreement.
Beyond the practical interest of this result for climate modelling, it supports the idea that, to a certain extent, climate can be characterized with macroscale features with no need to compute the underlying microscale dynamics.
\end{abstract}

\section{Introduction}

The Earth receives a certain amount of energy from the Sun, in the form of visible light, which it has to radiate back to space, in the form of infrared light, to maintain a steady state. Most Earth System processes, including weather and climate, can be regarded as little more than steps in this process of energy conversion from one form to the other, going through various other forms of energy (potential energy, kinetic energy, heat...). Although in each of these steps the overall quantity of energy has to be conserved, its quality can vary \cite{Peixoto1991, PeixotoPOC}. The field of physics that deals with such energy conversions in different forms and the corresponding evolution of its quality is called thermodynamics, and the quality of energy is measured by a state function called \emph{entropy}. 
The constraint of conservation of the exchanged quantity of energy is well-known and commonly used in climate sciences. Less attention is paid to the additional information one can gain by monitoringcreation and exchanges of entropy, either as a diagnostic tool or as a prediction principle relying on an extremum property. This information can be discarded with little consequences for some processes for which physical laws naturally allow for a macroscopic description (and for the subsequent computation of the exchanged energy due to the process), but in some other cases it might bring in a fundamental constraint that should not be ignored.

A typical example is atmospheric heat transport: considering the observational fact that the net top-of-atmosphere radiative budget is positive in the Tropics and negative near the Poles, one has to invoke a poleward atmospheric transport to ensure global energy balance in the steady state (see e.g. \cite{Trenberth1994} for a quantitative discussion based on observations and \cite{Lucarini2010c} for a modelling counterpart). An important difficulty comes from the turbulent nature of the laws of atmospheric motion and the lack of a satisfactory theory of turbulence. We are thus forced to integrate numerically the equations of motion at a very high computational cost. On the other hand, the laws of radiation allow for a macroscopic description more easily. To cope with this difficulty, the options up to now have been either to deal with the overdetailed (from the minimalist standpoint we are considering) microscopic equations of motion of the atmosphere, or to use empirical parameterizations.

An alternative was suggested by Paltridge \cite{Paltridge1975, Paltridge1978} (see \cite{OBrien1995} for a nice reformulation). He was looking for a large-scale description of heat transport and tested a variational principle stating that atmospheric heat transport adjusts so as to maximize the production of entropy. Several climate studies based on similar models have been published since this seminal work: \cite{Grassl1981,Gerard1990,Wyant1988,Lorenz2001} for zonally-averaged climate, \cite{Ozawa1997,Pujol2002} for vertical convection.
The Maximum Entropy Production can be interpreted as a maximum energy transport efficiency requirement. The general idea was first expressed by Lorenz \cite{Lorenz1960}. See for instance \cite{PeixotoPOC,Lin1982,Vos1993,Lucarini2009,Lucarini2010a} and references therein for discussions in terms of available potential energy and Carnot engines, and \cite{Pauluis2002a,Pauluis2002b} for the competing effect of water vapour removal by the atmosphere.

The purpose of this paper is not to discuss the justification of the Maximum Entropy Production Principle. For a rather comprehensive review of different approaches and applications, the reader is refered to \cite{Martyushev2006} and \cite{kleidonlorenzbook}.
In this paper, we present a new formulation of Paltridge's model: we have developed a different approach for the treatment of radiation, which we believe to be more rigorous and physically sound. In particular we provide a derivation of the analytical expression of the radiative coefficients which are introduced as \emph{ad hoc} in Paltridge's model. We also avoid the questionable maximum convection hypothesis, critical in Paltridge's model \cite{OBrien1995,Pujol2000}, by applying the MEP principle on the vertical dimension as well, and discard Paltridge's cloud cover variable and oceanic transport for the sake of cogency. Formulated this way, the model is as free of tunable coefficient, empirical parameterization and spurious assumption as possible, therefore constituting a clean basis to evaluate the validity of the MEP hypothesis as applied to climate.
The only information needed \emph{a priori} to compute the coefficients are the surface albedo, the vertically-integrated water vapour and ozone density and the $CO_2$ concentration.
In particular, we are now able to run a sensitivity experiment with respect to the surface albedo parameter. Here, we investigate the effect of the surface albedo distribution corresponding to the presence of large ice-sheets in the Northern Hemisphere during the Last Glacial Maximum. 
The results for pre-industrial and LGM climates are compared with standard simulations with the IPSL\_CM4 atmosphere-ocean general circulation model using similar boundary conditions.

\section{Model description}

\begin{figure*}[hp]
\centering
\includegraphics[width=0.5\linewidth]{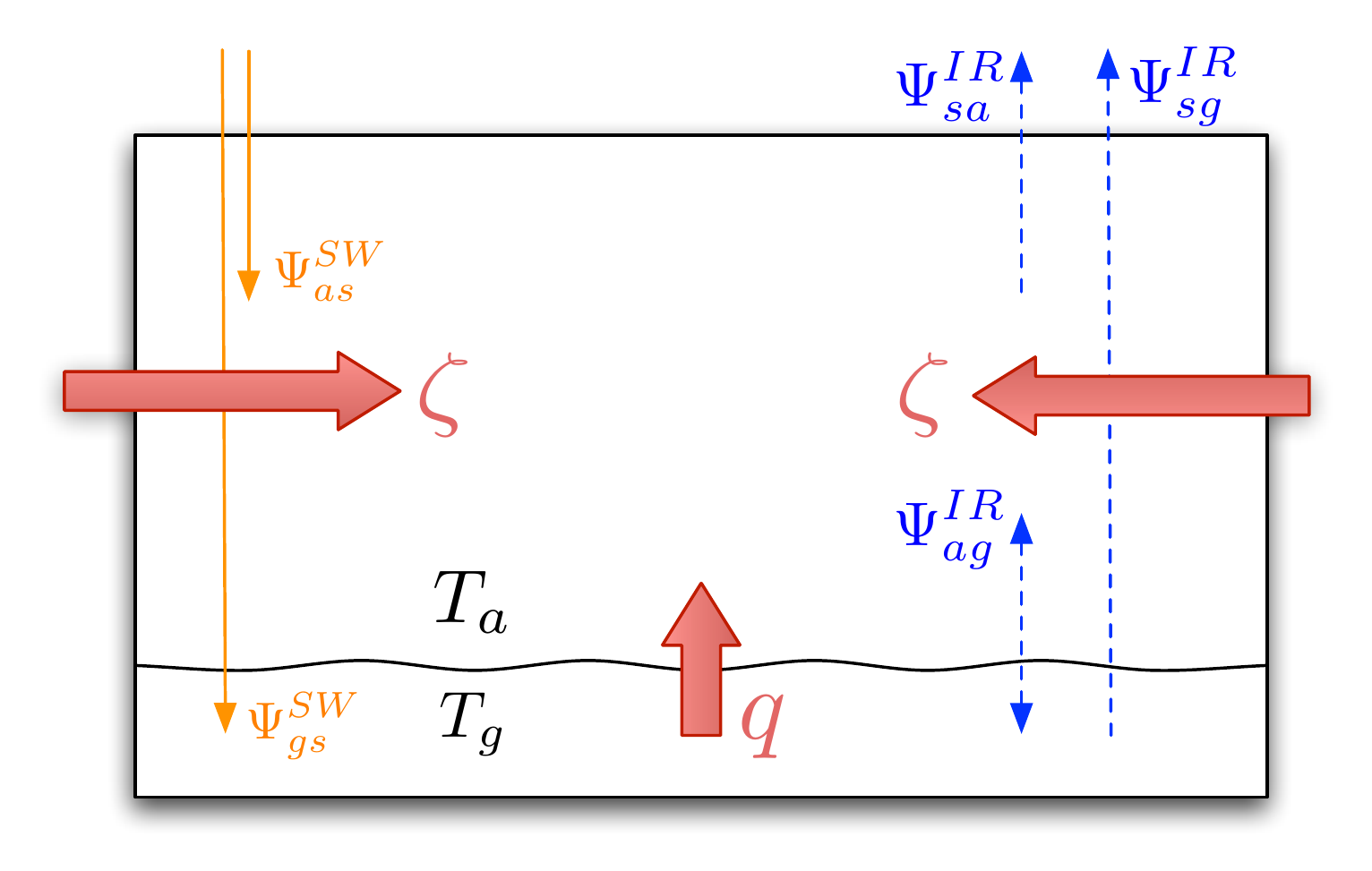}
\caption{Energy exchanges in one model box : incoming solar energy (thin solid lines) absorbed by atmosphere $\Psi_{as}^{SW}$ and ground $\Psi_{gs}^{SW}$, longwave radiative exchanges (dashed) ; exchange between the surface and the atmosphere $\Psi_{ga}^{IR}$, cooling to space of the surface $\Psi_{sg}^{IR}$ and of the atmosphere $\Psi_{sa}^{IR}$) and large scale heat transport convergence ($\zeta$) and surface sensible heat ($q$) (thick solid arrows).}\label{modelboxfig}
\end{figure*}

We use a purely thermodynamic two dimensional (horizontal) energy balance model. Only energy transfers are represented. The exchanges of mass, momentum, angular momentum, etc are discarded here. Each grid cell consists of two layers: one atmospheric layer (at temperature $T_a$) exchanging energy with the surface, space and adjacents cells in the form of longwave and shortwave radiation, surface heat flux, and atmospheric heat transport, and one surface layer (at temperature $T_g$), only exchanging energy with the overlying atmospheric layer and with space, in the form of shortwave and longwave radiation and surface sensible heat (see figure \ref{modelboxfig}).
We note $A_{ij}$ the area of cell $(i,j)$, $1\leq i \leq N_{lat}, 1\leq j \leq N_{lon}$. All the simulations presented in this paper are carried out on a regular grid with $N_{lat}=72$ and $N_{lon}=96$ for easier comparison with the GCM results using the same grid.

%Figure1

\subsection{Net-exchange formulation for longwave radiative transfer}

Radiative exchanges are described in the \emph{Net Exchange Formulation} (NEF) formalism - originally introduced by \cite{Green1967} (see \cite{Dufresne2005,Eymet:2004p215} for details). In the NEF, radiative fluxes are replaced by energy exchange rates between two layers as basic objects describing the radiative transfer. Instead of solving the radiative transfer equation, we compute the net energy exchanged per time unit between layers $i$ and $j$ (resp. layer $i$ and boundary surface $j$ or boundary surfaces $i$ and $j$) at frequency $\nu$:
\begin{equation}
\begin{split}
\psi_{ij}^\nu=\int_{\Sigma_i}\int_{\Sigma_j}\int_{\Gamma_{ij}}& dP_idP_jd\gamma \\
&\alpha_\nu(P_i,\gamma)\alpha_\nu(P_j,\gamma)[B_\nu(T_j)-B_\nu(T_i)]\tau_\gamma^\nu
\end{split}
\end{equation}

where $\Sigma_i$ represents the volume of layer $i$ (resp. the area of boundary surface $i$), $\Gamma_{ij}$ the set of optical paths from one point of layer $i$ to one point of layer $j$, $\alpha_\nu(P_i,\gamma)$ is $1$ if $i$ is a boundary surface and the absorption coefficient $k_\nu(P_i,\gamma)$ otherwise, $\tau_\gamma^\nu=\exp\left(-\int_\gamma k_\nu(s)ds\right)$ is the transmitivity of the path $\gamma$ and finally $B_\nu(T)$ is the Planck function. Depending on the nature of $\Sigma_i$ (surface or volume), $dP_i$ is either an infinitesimal surface or volume element. Clearly $\psi_{ij}^\nu=-\psi_{ji}^\nu$.

The NEF was originally used in engineering rather than atmospheric sciences. In the few existing cases of atmospheric transfer studies, the main point is that NEF allows for a fine analysis of the greenhouse effect and how it evolves when one parameter changes: it is clear from the $(\psi_{ij}^\nu)$ matrix which exchanges prevail in the atmosphere (generally, these are the exchanges between neighbouring layers, between the surface and any layer and the cooling to space of any layer and the surface).
This description contains \emph{more information} on the radiative transfer than the traditional flux approach insofar as one not only knows how much energy is radiated through a surface but also where the radiation originates from (computationally, for a $n$ layer atmosphere, NEF requires $O(n^2)$ numbers while the flux approach only needs $O(n)$).
Other major advantages of this formulation are the handiness it provides in coupling radiative exchanges to other modes of energy transfer, and the fact that it automatically enforces physical laws such as energy conservation or the direction of energy exchanges depending on the temperature gradient, whatever numerical approximations are made. 
Clearly these points make it suitable for simple thermodynamic models such as ours.

In the very simple case of our two level cell, we only have to compute the infrared exchange between the surface and the atmosphere $\Psi_{ga}^{IR}$, the surface and space $\Psi_{gs}^{IR}$ and the atmosphere and space $\Psi_{as}^{IR}$.
Applying the diffusive approximation ($\mu=1/1.66$ \cite{Elsasser1942}, justified in \cite{Li2000}), angular and vertical integration ($H$ being the height of the atmosphere) yields, for instance, for $\Psi_{ag}^\nu$:
\begin{equation}
\Psi_{ag}^\nu=\Sigma_g \left(1-e^{-\frac{1}{\mu}\int_0^H k_\nu(z)dz }\right)(B_\nu(T_g)-B_\nu(T_a))
\end{equation}

From now on, let us consider the exchanged energy per surface area, keeping the same symbols $\Psi_{ij}$.
Integrating over the whole spectrum, the equations for the exchanged energy can be recast into the following form:
\begin{eqnarray}\label{psilweqn}
\Psi_{ag}^{IR}&=&t_a (T_g) \sigma T_g^4-t_a(T_a)\sigma T_a^4\\
\Psi_{sa}^{IR}&=&t_a (T_a) \sigma T_a^4\\
\Psi_{sg}^{IR}&=&\left(1-\frac{t_a (T_g)}{\mu}\right)\sigma T_g^4
\end{eqnarray}
where $t_a(T) = \mu \left(1-\displaystyle \int_0^{+\infty} \frac{B_\nu(T)}{\sigma  T^4} \tau_\nu d\nu \right)$ represents the emissivity of the atmosphere and $\tau_\nu=\exp\left(-\frac{1}{\mu}\int_0^H k_\nu(z)dz\right)$.
Of course the transmission functions of the atmosphere $\tau_\nu$ depend on the vertical profiles of absorbing gases, pressure and temperature.
In order to preserve the simplicity of the analytical expressions and to allow for comparison with Paltridge, we impose here the heavy constraint that the coefficients $t_a$ are constant in each cell.
Given a characteristic temperature for the local atmosphere and a humidity profile, we can evaluate these coefficients using Goody's statistical model \cite{Goody1952} and the spectral data from  \cite{Rodgers1966}. The temperature and humidity profiles used here come from a linear interpolation of McClatchey's reference atmosphere profiles \cite{McClatchey:1972p1301}. Hence we prescribe here the $t_a$ coefficients as functions of latitude only. 
In fact we do not need the full profiles here but only the vertically integrated values. Sensitivity tests were carried out to assess the dependence of the $t_a$ coefficient on the water vapour amount and characteristic temperature and its impact on the MEP climate (see section \ref{tasensitivitysection}).

\subsection{Shortwave radiation parameterization}

At latitude $\phi$, the annual mean insolation is given by
\begin{equation}
\begin{split}
W_A=&\frac{S}{2\pi^2 \sqrt{1-e^2}}\\
&\int_0^{2\pi } \sqrt{1-(\sin \phi \cos \epsilon - \cos \phi \sin \epsilon \sin u)^2 du}
\end{split}
\end{equation}

where $S$ is the solar constant, $e$ the eccentricity of the Earth's orbit and $\epsilon$ the obliquity.
The incoming top-of-atmosphere solar radiation in each cell is given by the latitudinal mean of $W_A$ over the cell, which we write $\xi \times S$, $\xi$ thus representing the annual mean of the cosine of the zenith angle of the Sun for the latitude zone (weighed by a distance factor) and is therefore a function of the orbital parameters.
In all the simulations considered here, we use the present-day values for the orbital parameters: $e\approx0.0167$ and $\epsilon\approx23.4\degre$. The Last Glacial Maximum orbital parameters do not differ much from those values.

Solar radiation exchange rates are parameterized as in \cite{Lacis1974}. With their notations, let us define
\begin{eqnarray}
\bar{s_a}(R_g)&=&0.353+\frac{0.647-\bar{R_r}(\xi)-A_{oz}(Mu_{O_3})}{1-\bar{\bar{R_r^*}}R_g }\\
s_a&=&A_{wv}(M\tilde{u})\\
s_a^*&=&A_{wv}\left(\left(M+\frac{5}{3}\right)\tilde{u}\right)-A_{wv}(M\tilde{u})
\end{eqnarray}

where $R_g$ is the surface albedo, $\tilde{u}$ (resp. $u_{O_3}$) the column-integrated water vapour (resp. ozone) concentration. $M$ is the magnification factor accounting for the slant path, $A_{wv}$ (resp. $A_{oz}$) are numeric functions corresponding to water-vapour (resp. ozone) absorption, and $\bar{R_r}(\xi)$ and $\bar{\bar{R_r^*}}$ account for Rayleigh scattering in the atmosphere (see \cite{Lacis1974} or \cite{Stephens1984} for details).
Incoming solar radiation at surface and solar radiation absorbed in the atmosphere are then expressed in the simple form:
\begin{eqnarray}
\Psi_{gs}^{SW}&=&(\bar{s_a}(R_g)-s_a)(1-R_g)\xi S \label{psisweqn1}\\
\Psi_{as}^{SW}&=&(s_a+R_gs_a^*)\xi S\label{psisweqn}
\end{eqnarray}

The coefficient $s_a$ can be interpreted as the proportion of solar radiation directly absorbed in the atmosphere while $s_a^*$ accounts for absorption after reflection at surface and $\bar{s_a}$ represents the proportion of solar radiation absorbed by the ground.

To compute the value of the three coefficients $s_a,\bar{s_a},s_a^*$, we use humidity and ozone profiles from McClatchey \cite{McClatchey:1972p1301} again (as for the longwave coefficients, this is not a critical parameter here), with pressure scaling for water vapour.
Surface albedo is prescribed from IPSL simulations (see section \ref{albipslpar}).

\subsection{Entropy Production Maximization}\label{epmaxsection}

For each grid box, we can compute the parameters $(\xi,t_a,s_a,s_a^*,\bar{s_a},R_g)$ as explained above and radiative energy exchange rates are given by equations \ref{psilweqn} to \ref{psisweqn} as functions of these parameters and temperatures $T_a$ and $T_g$.

Since we are looking for steady states of the system, we impose the following energy balance constraints in each box:
\begin{eqnarray}\label{energybalanceeq1}
\Psi_{gs}^{SW}+\Psi_{as}^{SW}-\Psi_{sg}^{IR}-\Psi_{sa}^{IR}+\zeta&=&0\\
\Psi_{gs}^{SW}-\Psi_{ag}^{IR}-\Psi_{sg}^{IR}-q&=&0\label{energybalanceeq2}
\end{eqnarray}

where $q$ is the surface heat flux and $\zeta$ the horizontal convergence (see figure \ref{modelboxfig}). Formally these equations can be solved to express $T_a$ and $T_g$ as functions of the two unknown variables, $q$ and $\zeta$.

Similarly to the energy budget, we can write an entropy budget, separating the rate of change of the entropy of a parcel of the system into two contributions:
\begin{equation}
\frac{dS}{dt}=\frac{dS_e}{dt}+\sigma
\end{equation}
where $\dot{S_e}$ stands for the entropy exchanged with the surroundings per unit time, while $\sigma$ is the entropy created per unit time. The second law of thermodynamics states that $\sigma \geq 0$. In the steady-state, we have $\dot{S}=0$.

Assuming local thermodynamic equilibrium, the variation of the entropy for a parcel of the system exchanging an amount of heat $\delta Q$ is given by $dS=\delta Q/T$. Hence, for a single cell of the model, the entropy production rate (per unit surface area) associated to the horizontal convergence is given by $\frac{\zeta}{T_a}$ and the entropy production associated to the surface heat flux is $q\left(\frac{1}{T_a}-\frac{1}{T_g}\right)$.

We thus consider the total \emph{material entropy production rate}:
\begin{equation}
\begin{split}
&\sigma_M(\{q_{ij},\zeta_{ij}\})=\\
&\sum_{i=1}^{N_{lat}}\sum_{j=1}^{N_{lon}} \left(\frac{q_{ij}}{T_{a,ij}}-\frac{q_{ij}}{T_{g,ij}}+\frac{\zeta_{ij}}{T_{a,ij}}\right)A_{ij}
\end{split}\label{sigmadefeq}
\end{equation}

Let us search for the energy fluxes distribution maximizing this function subject to the following global constraint: the total heat transport divergence over the globe must vanish, \emph{i.e.}
\begin{equation}
\sum_{i=1}^{N_{lat}}\sum_{j=1}^{N_{lon}} A_{ij}\zeta_{ij}=0\label{constrainteq}
\end{equation}
We enforce this physical condition by introducing a Lagrange multiplier $\beta$ in the previous equation.

\begin{equation}
\begin{split}
&\sigma_M(\{q_{ij},\zeta_{ij}\},\beta)=\\
&\sum_{i=1}^{N_{lat}}\sum_{j=1}^{N_{lon}}A_{ij}\left(\frac{q_{ij}}{T_{a,ij}}-\frac{q_{ij}}{T_{g,ij}}+\frac{\zeta_{ij}}{T_{a,ij}}-\beta \zeta_{ij}\right)\\
\end{split}
\end{equation}

and we call \emph{Maximum Entropy Production} (MEP) state the solution $\left\{T_{a,ij}^*,T_{g,ij}^*,q_{ij}^*,\zeta_{ij}^*\right\}_{1\leq i\leq N_{lat}, 1 \leq j \leq N_{lon}}$ of the system (\ref{energybalanceeq1})-(\ref{energybalanceeq2}) maximizing the entropy production rate (\ref{sigmadefeq}) with the imposed constraint (\ref{constrainteq}). In practice, solving the system reduces to finding the maximum of a function of two variables in each box.
One could equivalently solve the optimization problem in terms of the temperature variables, replacing $q$ and $\zeta$ from equations \ref{energybalanceeq1}-\ref{energybalanceeq2} into $\sigma_M(T_a,T_g)$.

In the MEP procedure, it is only the entropy production rate due to turbulent processes which is maximized, motivated by the supposed connection of the MEP hypothesis with principles of maximum statistical entropy. The entropy production rate due to radiative processes, however, can provide interesting information as a diagnostic tool (see section \ref{entropbudgetsection}).

\subsection{IPSL simulations description and albedo forcing}\label{albipslpar}

The IPSL model is a fully coupled Ocean-Atmosphere General Circulation Model (OAGCM). The version used here is IPSL\_CM4 \cite{Marti2010} i.e. the version which has been used for the CMIP3/IPCC AR4 exercise. The model is composed of the LMDZ atmosphere model \cite{Hourdin2006} run at resolution 96x72x19 (longitude-latitude-vertical levels), the ORCA2 ocean model \cite{Madec1997}, the LIM2 see-ice model \cite{Fichefet1997,Fichefet1999}, all coupled via the OASIS coupler \cite{Valcke2006}.

The present work uses two simulations performed with this model : (1) a pre-industrial simulation, for which the model is forced with constant, pre-industrial greenhouse gas concentration ($CO_2 = 280$ ppm, $CH_4 = 760$ ppb, $N_2O = 270$ ppb), insolation, coastlines, topography, land-ice extent and (2) a pseudo LGM run. Since the simple climate model described above does not account for topography, nor for $CO_2$ variations, we have chosen to use an OAGCM simulation in which the extent of the LGM ice-sheets is taken into account but not their height. The ice-sheet extent is prescribed from the Peltier ICE-5G reconstruction \cite{Peltier2004}. Greenhouse gases atmospheric concentrations are maintained at their pre-industrial level, as well as orbital parameters and topography.
Note that none of the simulations uses an aerosol forcing, which explains that the simulated pre-industrial climate is slightly warmer than the standard IPCC simulation.

\begin{figure*}[hp]
\centering
\includegraphics[width=\linewidth]{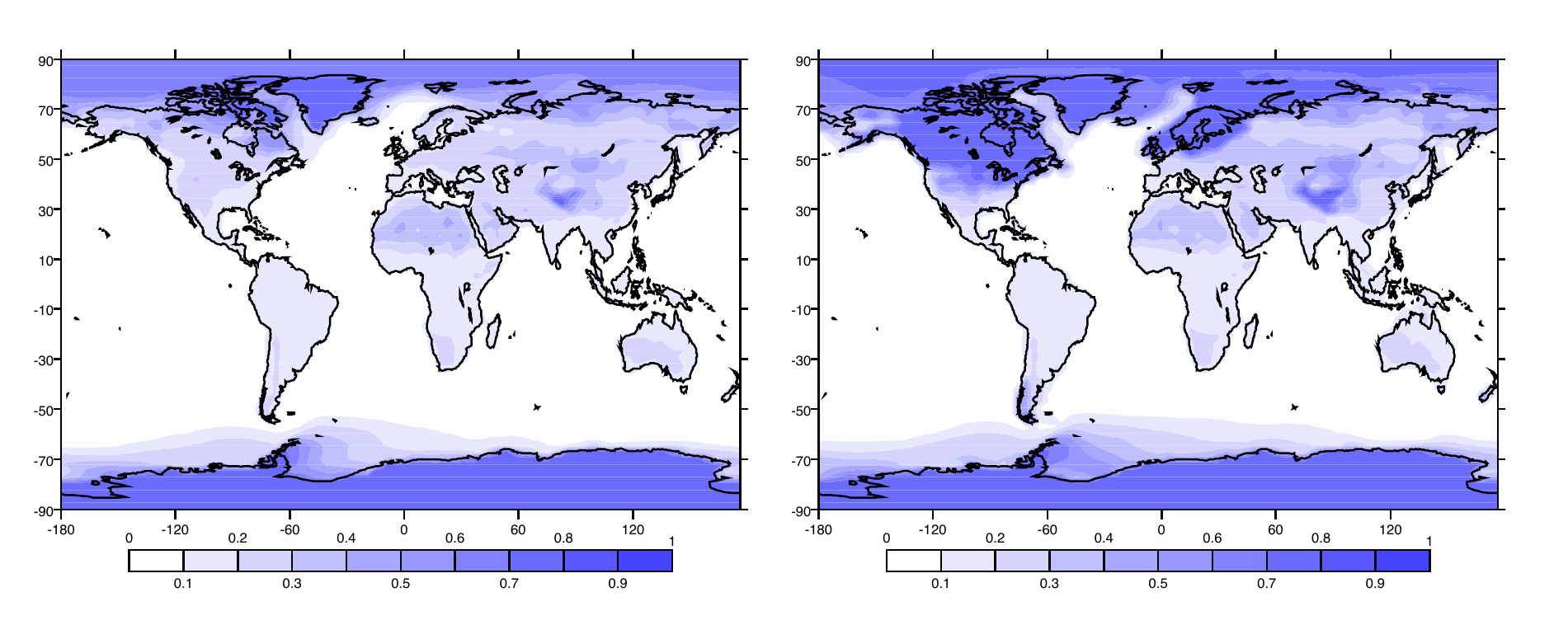}
\caption{Surface albedo forcing for the MEP model, computed from the output of the IPSL\_CM4 simulations. Left : annual mean surface albedo values from the pre-industrial simulation with the IPSL model. Right : annual mean surface albedo values from the LGM ice extent simulation with the IPSL model. The surface albedo is higher over continents than over the open ocean, and it is higher in high-latitude areas than in low-latitude areas because of the presence of snow and ice. The main difference between pre-industrial and LGM surface albedo distributions is the presence of a large ice-sheet over North America and Scandinavia during the LGM.}\label{surfalbedomap}
\end{figure*}

For each of these simulations, the annual mean surface albedo is computed (figure \ref{surfalbedomap}) and used as an input for the corresponding MEP simulation.

%Figure2

Hence the boundary counditions for the IPSL\_CM4 simulations correspond as much as possible to those for the MEP simulations.

\section{Results and discussion}

\subsection{Present day climate}\label{presentresultssection}

\begin{figure*}[hp]
\centering
\includegraphics[width=\linewidth]{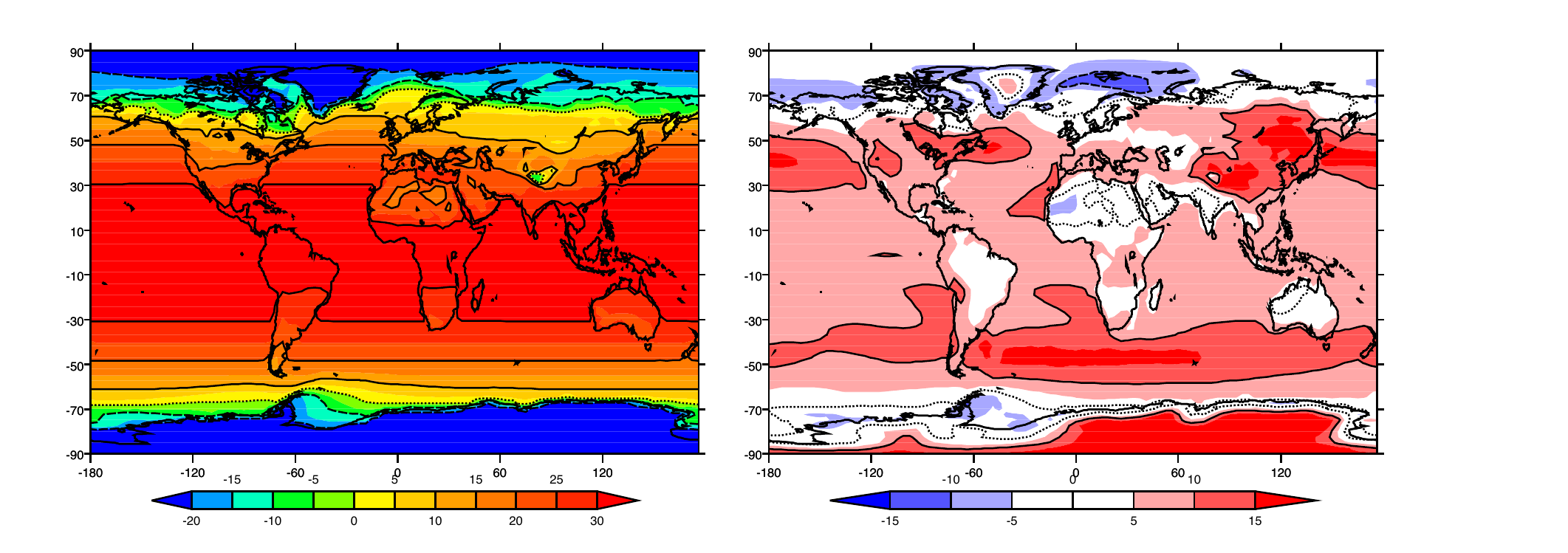}
\caption{Results for surface temperature. Left: Surface temperature at MEP state for present conditions, in $\degre$C. Right: Surface temperature difference between the MEP state and the IPSL simulation. Contour lines interval is $10 \degre$C, positive contours are drawn in solid lines, negative contours in dashed lines and the null contour as a dotted line. 
}\label{tgpstmap}
\end{figure*}

Figure \ref{tgpstmap} shows the surface temperature at MEP state for present albedo forcing and a comparison with a control simulation from the IPSL model. 

%Figure3

The global mean temperature simulated by our simple MEP model is $23\degre$C, which is $7\degre$C warmer than the IPSL model results. The reason for this strong bias is probably the lack of any clouds in the model. 
Another reason could be the crude representation of the vertical structure of the atmosphere in the model and the subsequent flaws in vertical radiative transfer calculations. Comparison with the IPSL model reveals that some of the regions of largest discrepancies are regions of high elevation (Antarctica, Tibetan plateau, and to a smaller extent Greenland, cf Fig. \ref{tgpstmap}). Indeed, topography effects are not included in our model. We show in section \ref{cloudssection} that the combined effects of clouds and elevation account for the bulk of the global difference with the IPSL model.

In spite of this global bias, considering the simplicity of the model and the absence of any degree of freedom to adjust the results, the model gives a strikingly accurate view of the global picture of the climate. Some important specific features, like the presence of the main deserts (Sahara, Arabian peninsula, Kalahari, Australia,..), are already visible taking into account only the albedo effect.
Since the only non zonal parameter in the model is the albedo, and its value is roughly uniform over the oceans, we were expecting the surface temperature to be zonally homogeneous over the oceans, which is indeed the case. The positive anomaly of temperature over mid-latitudes regions  (compared to the IPSL results, Fig. \ref{tgpstmap}, right) is certainly imputable to the important cloud cover over those areas, lacking in the model.

%Figure4

\begin{figure*}[hp]
\centering
\includegraphics[width=\linewidth]{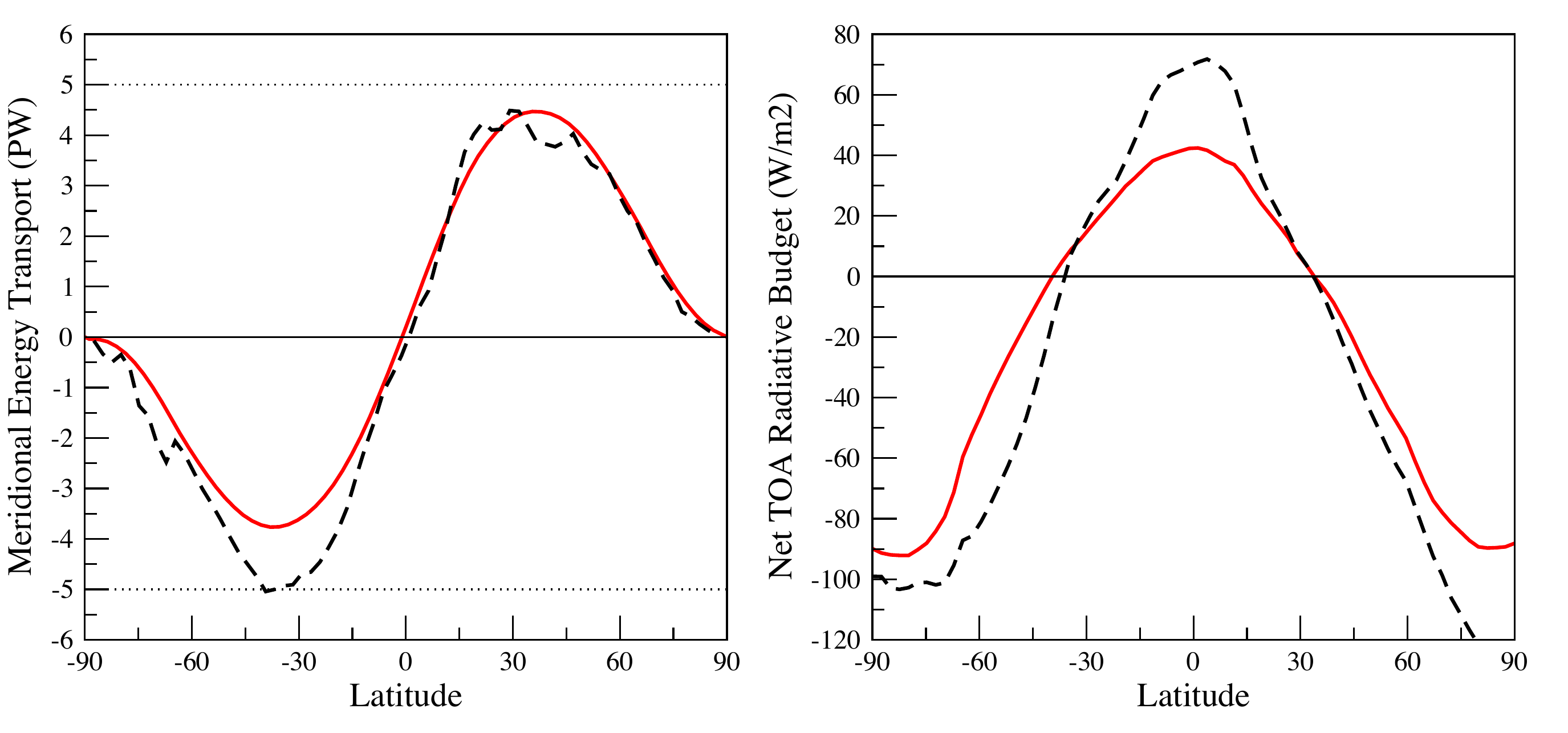}
\caption{Left: Total Meridional energy transport predicted by MEP for present conditions (solid) and for the IPSL model (dashed), in PW. 
Right: Net radiative budget at the top of atmosphere as a function of latitude (zonally averaged) for the MEP state (solid) and for the IPSL model (dashed) at present conditions, in $W.m^{-2}$. 
}\label{transportplot}
\end{figure*}

Figure \ref{transportplot} provides the curves of zonally averaged atmospheric heat transport and net top of atmosphere radiative budget. The net top of atmosphere radiation budget switches from positive in the Tropics to negative near the Poles at roughly 40$\degre$S and 35$\degre$N, while the overall shape of atmospheric heat transport qualitatively fits the IPSL model results. It peaks approximatively at 35$\degre$S and 40$\degre$N with respective values of 3.8 PW and 4.4 PW. The location of these maximum values is consistent with \cite{Trenberth2001} but the absolute value is slightly underestimated here: \cite{Trenberth2001} gives the peak transport around 6 PW. The absence of any oceanic heat transport in our model prevents us to push the comparison forward. 
Comparison with the atmospheric transport from the IPSL simulation reveals that the MEP transport is indeed a little low, especially in the Southern Hemisphere.

%Figure5

\begin{figure*}[hp]
\centering
\includegraphics[width=\linewidth]{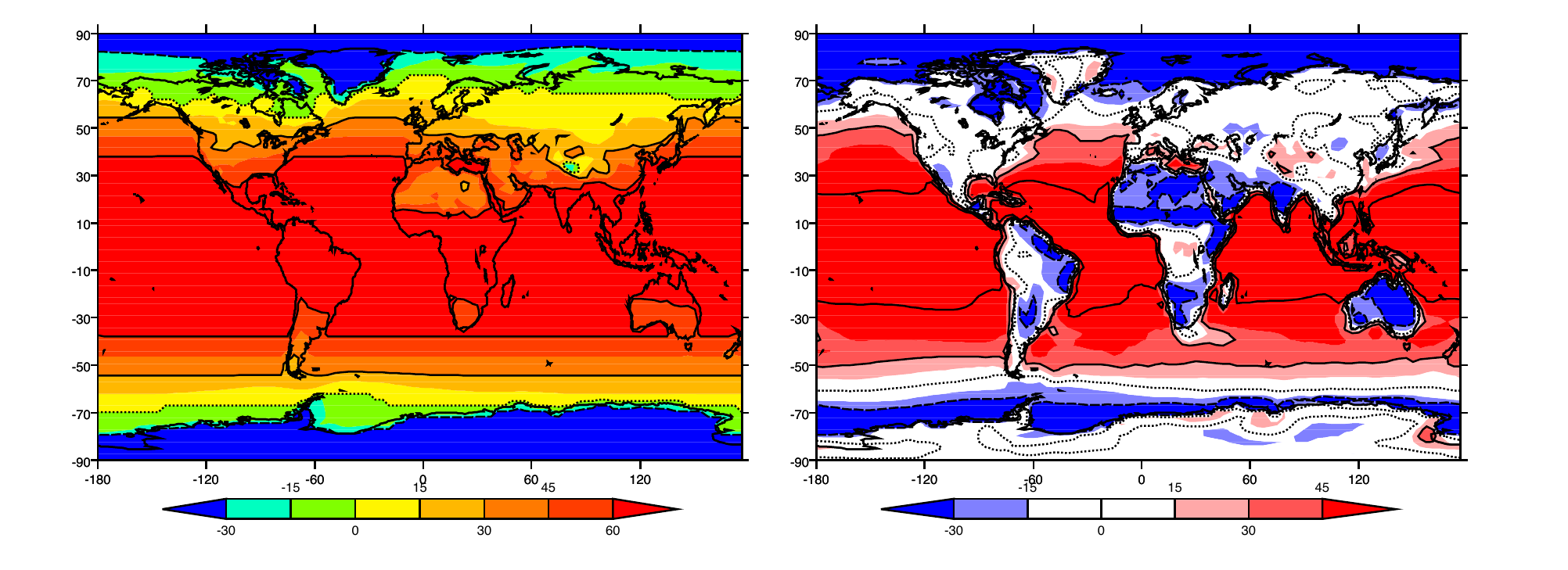}
\caption{Left: Surface heat flux predicted by MEP for present conditions, in $W.m^{-2}$. The mean value is $55 W.m^{-2}$. Right: Difference between the surface energy flux at MEP state and the surface sensible heat flux for the IPSL simulation for present conditions. Contour lines interval is $30 W.m^{-2}$, positive contours are drawn in solid lines, negative contours in dashed lines and the null contour as a dotted line.}\label{sfcheatmap}
\end{figure*}

It might also explain why surface energy fluxes are underestimated here: since all the heat flowing laterally is assumed to come from the atmosphere, the need for surface heat flux is reduced. Indeed, in the MEP state, the global mean surface flux is only $55 \mbox{W.m}^{-2}$ (figure \ref{sfcheatmap}), compared to $97 \mbox{W.m}^{-2}$ \cite{Trenberth2009}. The maximum value of the surface energy flux is reached in the Tropics while it is negative in the polar regions. Comparison with the IPSL simulation (not shown) confirms that the largest discrepancy arises over the oceans, where the lack of oceanic transport in our MEP model is crucial.

These results can also be interpreted in a different manner, considering that the surface energy flux computed by MEP is actually closer to the sensible heat component rather than the sum of the sensible and latent heat flux. Formally, the physical nature of the surface flux is not specified in the equations of the model. Yet, we believe that since it appears in the entropy production rate as the product between the flux and the gradient of the inverse of temperature, and given that the sensible heat flux also depends directly on the temperature gradient at the surface but not the latent heat flux, it makes more sense to compare it to the sensible heat flux (figure \ref{sfcheatmap}). In that case, the global average surface flux is approximately twice the value given by \cite{Trenberth2009}. The fact that the global mean average surface flux lies in between the value of the global mean sensible heat flux and the sum of the sensible and latent heat flux can be seen as a trade-off between the need for a large surface flux to ensure the energy balance and the need for a large enough temperature gradient from the entropy production rate point of view.

As discussed in \cite{Kleidon2003}, since a MEP state is supposed to be the most efficient heat conducting state given the constraints, one should expect more gentle equator-pole temperature gradients in MEP states than with more conventional models. Here, adjusting the overall surface temperature field by setting $T_g' = T_g +(\langle T_g \rangle_{IPSL}-\langle T_g\rangle_{MEP})$ to eliminate the global bias, we observe that the Tropics are roughly at the same temperature in the IPSL model and the adjusted MEP model (actually slightly colder in the MEP model) while the mid-latitudes are substantially warmer (around 5 degrees) in the MEP state. However the temperature anomaly is not symmetric for the poles: the South Pole is colder by around $10\degre$C in the adjusted MEP model whereas the North Pole is warmer by around $5\degre$C. Therefore the equator-pole gradient is indeed reduced in the MEP state for the Northern Hemisphere, but not for the Southern Hemisphere, as expected from the relatively low poleward heat transport in the Southern Hemisphere (figure \ref{transportplot}).

\subsection{A representation of clouds in the MEP climate}\label{cloudssection}

As suggested in section \ref{presentresultssection}, the major part of the difference between the MEP pre-industrial climate and the corresponding IPSL simulation is likely to be due to the absence of clouds in the MEP model. Inclusion of fixed or variable clouds in MEP models is possible, but unavoidably requires parameterization and extra assumptions, which is not the philosophy of this study. Nevertheless, to evaluate the effect of clouds on the MEP model presented here, it is possible to modify the radiation equations (\ref{psisweqn1})-(\ref{psisweqn}) to include an \emph{ad hoc} cloud parameter $R_n$:
\begin{eqnarray}
\Psi_{gs}^{SW}&=&(\bar{s_a}(R_g)-s_a)(1-R_g)(1-R_n)\xi S\\
\Psi_{as}^{SW}&=&(s_a+R_gs_a^*)(1-R_n)\xi S
\end{eqnarray}

The parameter $R_n$ is computed in each grid box from the cloud radiative forcing at the top of the atmosphere in the corresponding area in the IPSL model.
On the global mean, the cloud radiative forcing at the top of the atmosphere in the IPSL model is roughly $-20 W.m^{-2}$ (in accordance with \cite{Kiehl1997}), which corresponds to a mean value of $R_n \approx 0.073$, but the geographical distribution of the radiative forcing is very far from being uniform.

%Figure6
 
\begin{figure*}[hp]
\centering
\includegraphics[width=0.5\linewidth]{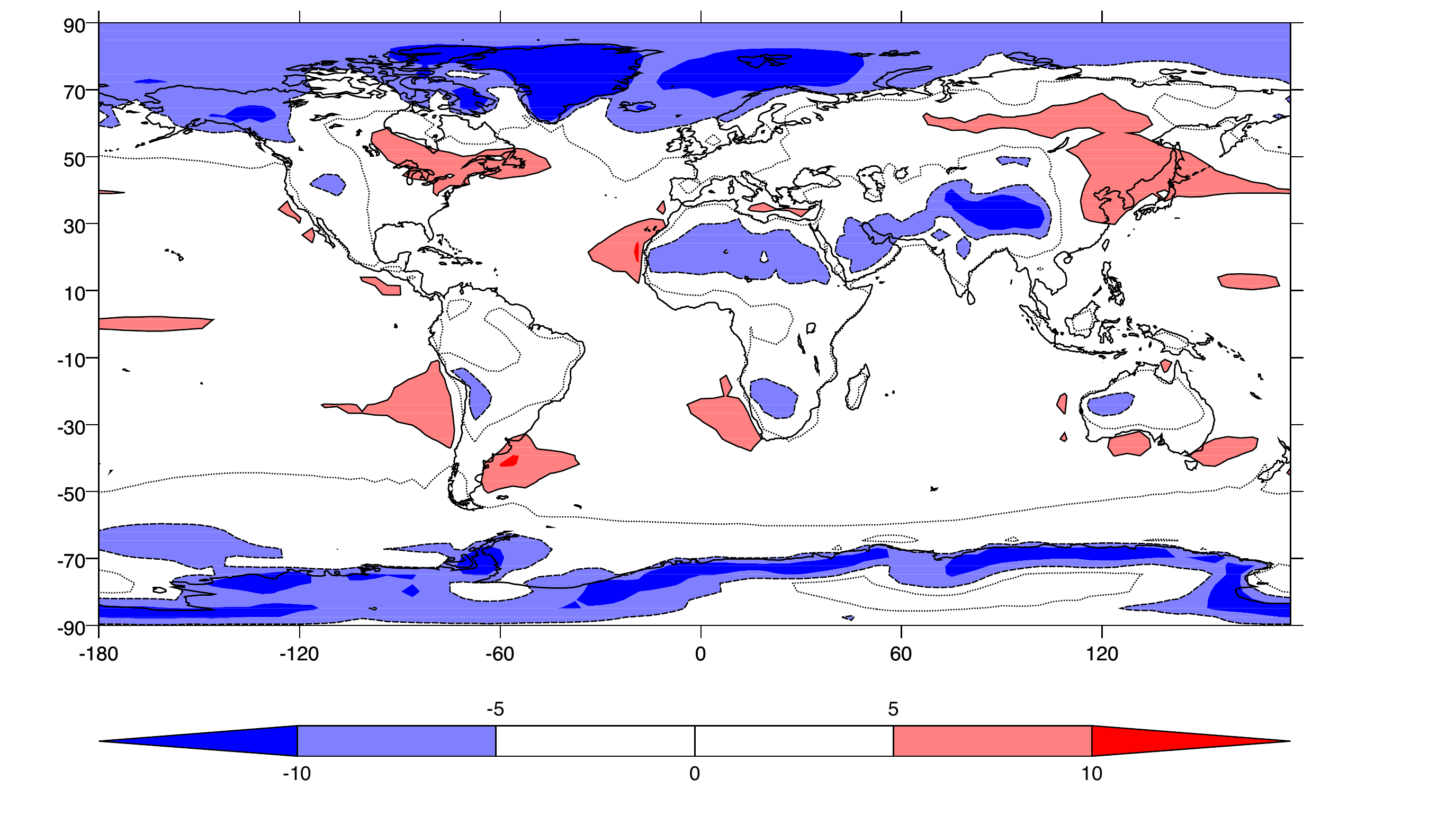}
\caption{Pre-industrial surface temperature difference between the MEP model with clouds, including a topography correction, and the IPSL model. Contour lines represent the -5 isoline (dashed), the null isoline (dotted) and the +5 isoline (solid). This figure is to be compared with figure \ref{tgpstmap}, right.}\label{figcloudstopo}
\end{figure*}

For pre-industrial climate, this parameterization of clouds in the MEP model reduces the temperature by 4.7$\degre$C on global average. Thus a 2.5$\degre$C difference remains with the corresponding IPSL run. The temperature anomaly map (not shown) reveals that this difference is mainly due to the effect of topography (Antarctica and Tibet are in particular very warm in the MEP model), not accounted for in the MEP model. As an approximate correction, we apply a constant $-7 K.km^{-1}$ lapse rate, using the elevation data from the IPSL run output. Figure \ref{figcloudstopo} shows the combined effect of clouds and topography on the MEP pre-industrial climate: the global mean difference with the IPSL pre-industrial run becomes 0.9$\degre$C, which is comparable to the uncertainty due to the water vapour content of the atmosphere (see section \ref{tasensitivitysection}). Thus a crude parameterization of the main components lacking in the rigorous version of the MEP model suffices to recover the correct global picture of present climate. Note in particular that the warm bands in the mid-latitudes in figure \ref{tgpstmap} have completely disappeared with the inclusion of clouds.

\subsection{Effect of the radiative parameters on the MEP state}\label{tasensitivitysection}

As mentioned above, the only parameters in the model are the radiative parameters $(t_a,s_a,s_a^*,\bar{s_a})$ and the surface albedo $R_g$ (forced in each box with the surface albedo from the IPSL run). Computation of the radiative parameters only requires the value of the column-integrated water vapour density $u_{H_2O}$, and a characteristic temperature $T$ for $t_a$.

In this study, we have fixed once and for all the value of the radiative parameters as a function of latitude only, using a linear interpolation of standard atmospheric profiles for $u_{H_2O}$ and $T$. To assess the sensitivity of the model with respect to this choice, we have also computed the MEP state for pre-industrial surface albedo using a constant value for $t_a$, independent of latitude. Three choices were made, corresponding to the three standard profiles: Sub-Arctic, Mid-Latitude and Tropical. Table \ref{tasensitivitytable} shows the global mean temperature resulting from these uniform $t_a$ choices.

%Table1

The values for the column-integrated water vapour density and characteristic temperature span a realistic interval: as a comparison, in the IPSL pre-industrial run, $u_{H_2O}$ ranges from around $0.25 g.cm^{-2}$ near the poles to around $5.5 g.cm^{-2}$ at the equator. Thus we can estimate the sensitivity of the model to realistic choices in $u_{H_2O}$ and $T$ to be on the order of 1K.

\subsection{Last Glacial Maximum}

%Figure7

\begin{figure*}[hp]
\centering
\includegraphics[width=\linewidth]{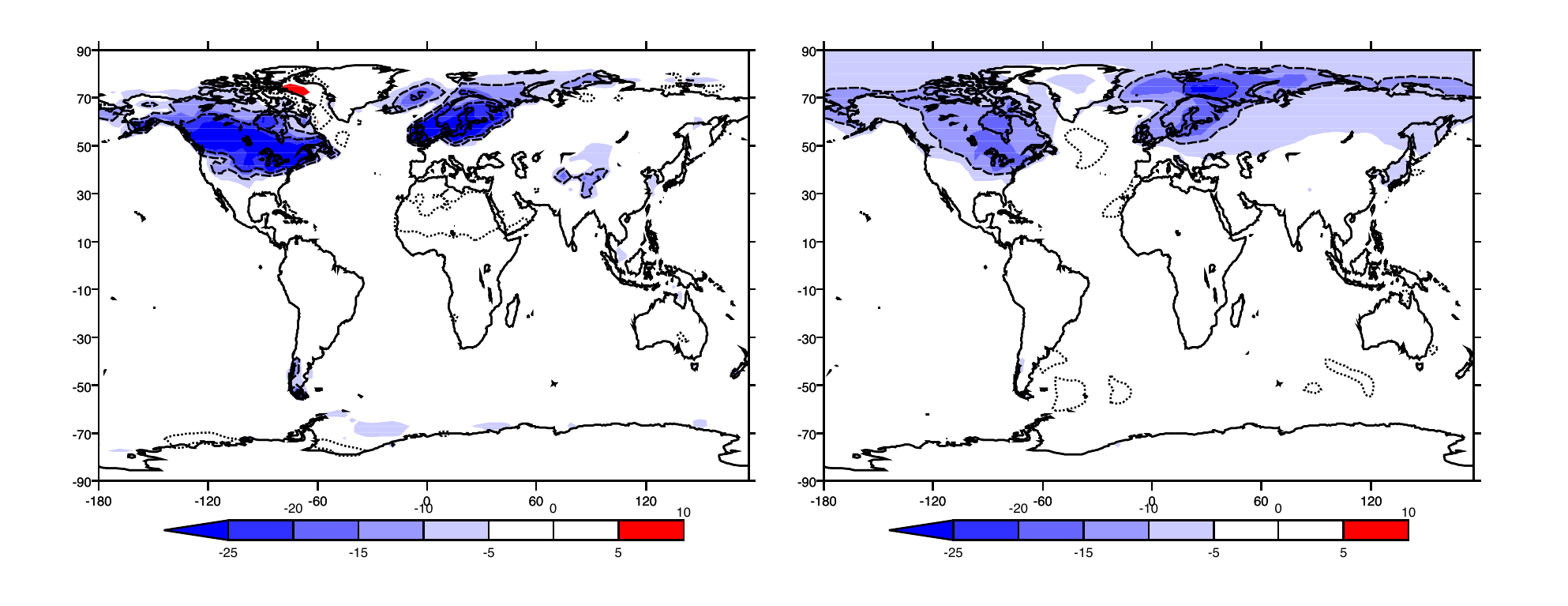}
\caption{Left: Surface temperature difference between Last Glacial Maximum conditions and present conditions at MEP states. Right: Surface temperature difference between Last Glacial Maximum conditions and present conditions for the IPSL model. Contour lines space is $10 \degre$C, positive contours are drawn in solid lines, negative contours in dashed lines and the null contour as a dotted line. The global mean LGM cooling is $-2\degre$C for the MEP model and $-2.5\degre$C for the IPSL model. }\label{tglgmmap}
\end{figure*}

Figure \ref{tglgmmap} shows the surface temperature difference between the Last Glacial Maximum and pre-industrial control run, at MEP and with the IPSL model. The global mean difference is $\approx 2$ K colder for the LGM at MEP, compared to $\approx 2.5$ K colder for the IPSL model. This remarkable overall agreement means that the sensitivity of the MEP model to surface albedo is comparable to that of the GCM.
The major part of the cooling occurs over Canada and Northern Europe for both models, but it is much stronger in the MEP model: between 20 and 30 K for these areas, compared to $\approx 15$K for the IPSL model. The cooling is also overestimated in Patagonia - by a factor 2 approximatively - and Tibet  - by a factor 3.
These extreme values are compensated for by a small underestimate of the cooling almost everywhere else: around $-0.5$ K for most of the oceans, South America, Africa, Australia, Antarctica and Southern and Eastern Asia vs $-1$ to $-2$ K for the IPSL model. This milder cooling becomes even more dramatic over Russia - $-2$ K for MEP vs $-6$ K for the IPSL model - and Greenland - $-1$ K vs $-5$ K. 
Surprisingly the MEP model predicts an intense warming - $\approx 8$ K - in the Baffin Bay. We suspect this warming to be due to seasonality effects of the albedo.

Beyond the general relative agreement, the MEP model response to surface albedo is a little too strong over land (Canada, Northern Europe, Patagonia and Tibet) but not over sea: the Barents sea is slightly cooler in the IPSL-LGM than in the MEP-LGM.

%Figure8

\begin{figure*}[hp]
\centering
\includegraphics[width=\linewidth]{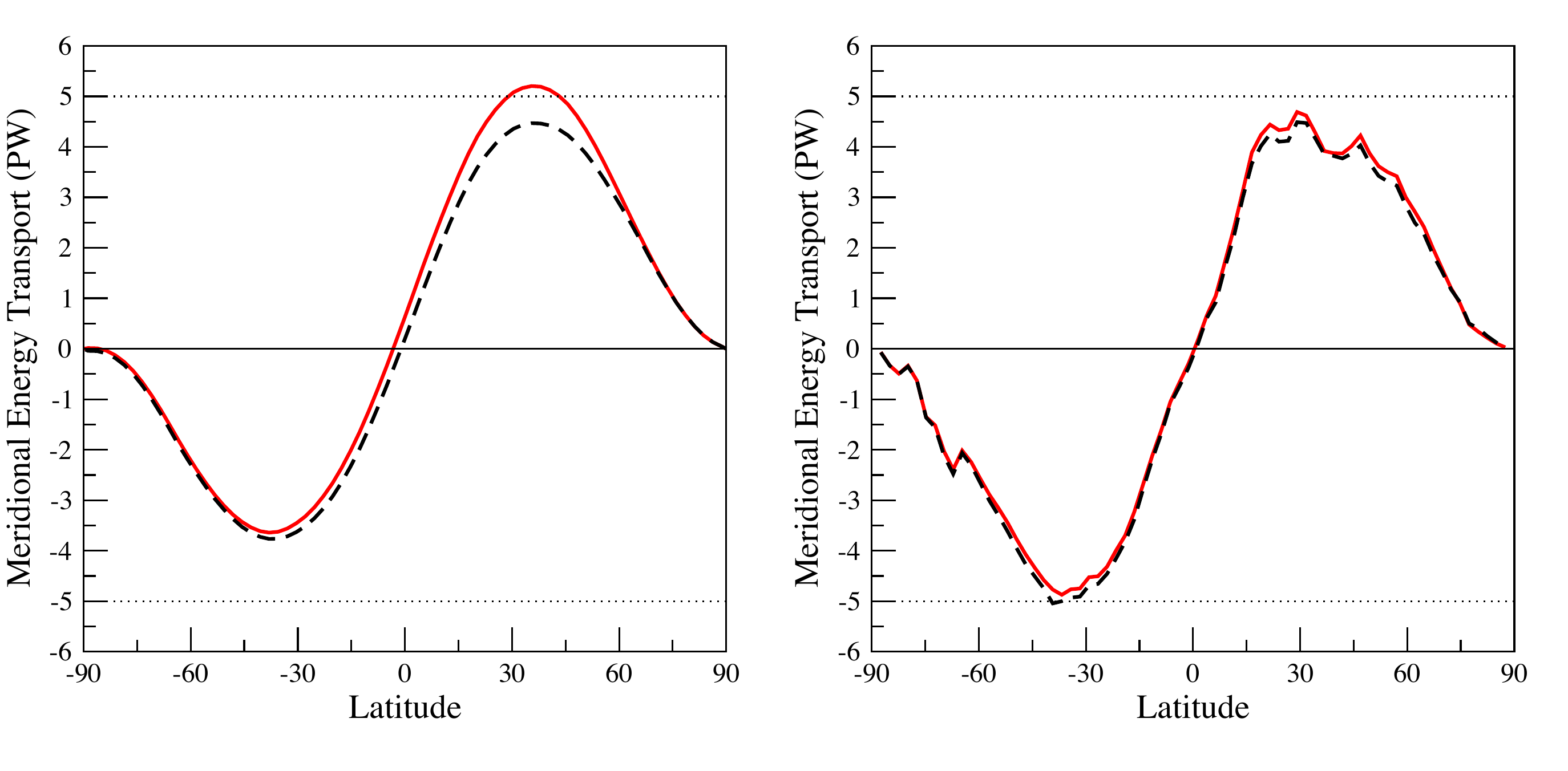}
\caption{Total Meridional energy transport predicted by MEP for LGM conditions (solid line) and present conditions (dashed line) at MEP state (left) and for the IPSL simulation (right), in PW.}\label{transportplotlgm}
\end{figure*}

Figure \ref{transportplotlgm} shows the meridional heat transport for the Last Glacial Maximum in MEP state. There is no major change of regime compared to present climate. Poleward transport is slightly more important (roughly $10\%$ more energy at the maximum) during the LGM, as can be expected from colder high latitudes.
This enhancement of the MEP-LGM poleward transport, larger than that obtained by the IPSL simulation, can probably be attributed to the exaggerated cooling of the Northern high latitudes in MEP-LGM.
During the Last Glacial Maximum, surface fluxes (not shown) become negative over ice-covered areas, the remaining of the globe showing little difference with present surface fluxes at MEP state.

The effect of clouds for the LGM at MEP state, with the cloud parameter being calculated from local cloud radiative forcing in the IPSL LGM simulation, is less than for pre-industrial climate. On the global mean, clouds cool the LGM climate by $4\degre$C in the MEP model. As a result, the sensitivity of the MEP model with clouds is reduced to $-1.3\degre$C, leading to a sensitivity difference with the IPSL model of $1.2\degre$ C, as compared to $0.5\degre$C for the model without clouds. Topography has no influence on the LGM-PI surface temperature difference as it is the same for both periods in both models.

Since the radiative coefficients are set once and for all in the MEP model, there is no representation of the water-vapour feedback, contrary to the IPSL model. In the IPSL model, the globally averaged water-vapour density changes from $2.86 g.cm^{-2}$ for pre-industrial climate to $2.50 g.cm^{-2}$ at the LGM, corresponding to a variation in $t_a$ of around $0.003$. From section \ref{tasensitivitysection}, one can estimate that the effect of water-vapour feedback on the MEP-LGM climate must be on the order of a few tenth of degrees.

\subsection{Entropy budgets}\label{entropbudgetsection}

We can decompose the total entropy production rate of section \ref{epmaxsection} into contributions from various processes as follows:
\begin{equation}
\sigma=\sigma_{SW}^{atm}+\sigma_{SW}^{sfc}+\sigma_{LW}^{atm-sfc}+\sigma_M
\end{equation}
where $\sigma_{SW}^{atm},\sigma_{SW}^{sfc},\sigma_{LW}^{atm-sfc}$ are the entropy production rates due to, respectively, absorption of solar radiation in the atmosphere, absorption of solar radiation at the surface and surface-atmosphere longwave interaction:
\begin{eqnarray}
\sigma_{SW}^{atm}&=&\Psi_{as}^{SW}\left(\frac{1}{T_a}-\frac{1}{T_{sun}}\right)\\
\sigma_{SW}^{sfc}&=&\Psi_{gs}^{SW}\left(\frac{1}{T_g}-\frac{1}{T_{sun}}\right)\\
\sigma_{LW}^{atm-sfc}&=&\Psi_{ag}^{IR}\left(\frac{1}{T_a}-\frac{1}{T_g}\right)\\
\end{eqnarray}

The entropy production rate of a thermodynamic process can be interpreted as a measure of its irreversibility. Therefore, the inspection of the relative values of the different terms above, gives valuable insight with respect to the role played by the different processes in driving the Earth out-of-equilibrium.

The expressions for the contributions to entropy production involving interaction between radiation and matter are only approximate, as noticed by \cite{Essex1984b}, but we shall not delve further into this here, both for simplicity and to compare with published results. The reader is referred to \cite{Stephens1993} or \cite{Wu2010}, for example, for computations taking into account the correct entropy of the radiation field. Nevertheless, the nonlocality of the NEF is already an improvement, in principle, as compared to the usual flux formulation, which is a local expression of radiative transfer \cite{Essex1984a}.

%Table2

Table \ref{entropybudgettable} shows the entropy budget of the Earth under pre-industrial and Last Glacial Maximum conditions at the MEP state. These values are compared with estimations for present climate from observations \cite{Peixoto1991}, from the intermediate complexity model PLASIM \cite{Fraedrich2008} and from the Hadley Centre GCM \cite{Pascale2009}.
In both cases, the entropy production due to absorption of solar radiation by the atmosphere is a little lower than other estimates. More importantly, the entropy production due to absorption of solar radiation at the surface is significantly higher than other estimates: this is due to the high solar absorption at the surface (in the absence of clouds) which is not compensated for by a warmer surface. The entropy production due to longwave interaction between the surface and the atmosphere (the greenhouse effect) is comparable to the values for PLASIM and HadCM3, and the material entropy production is considerably lower than other estimates. This last point is explained by the fact that latent heat fluxes are not explicitly represented in the MEP model, while they account for the major part of the material entropy production in all the other studies.
Note that the total entropy production at MEP state is slightly lower for the LGM as compared to the pre-industrial, in spite of a higher material entropy production.

\section{Conclusion}

The aim of this study was to provide a physically reliable, more consistent and more easily generalized version of the Paltridge model. The radiative part has been entirely reformulated using the \emph{Net-Exchange Formulation} in which the physical meaning of all the coefficients is clearly identified. Therefore the calculation of these coefficients, to the desired degree of approximation, only involves standard radiative data. Besides, all the energy fluxes other than radiation (horizontal convergence and surface fluxes) are subjected to the same principle: Maximum Entropy Production.

We have shown here that a thermodynamic model based only on a simple but robust radiative scheme and the principle of maximum entropy production yields results comparable to those of the IPSL model, for surface temperature, for both the pre-industrial and the Last Glacial Maximum climates, with a negligible computational cost. Obviously, the description adopted in the MEP model discards an important number of climate features. The inclusion of some key processes, like the seasonal or the hydrological cycles, in this thermodynamic framework would be of great interest. But the model in its present form already serves the purpose of providing a compromise between a minimal model with very little \emph{a priori} data and information needed, and acceptable efficiency.

In particular, one of the strongest points in the model is certainly that it does not include any adjustable parameter. This ability to get rid of the usual, varyingly important, parameter calibration makes it a good candidate to investigate climates where little is known or where some phenomena are likely to be different from the usual parameterization validity range, like for instance climates of other planets, inside or outside the solar system, or paleoclimates.

Further developments also include coupling the model with a state-of-the-art radiative code to come up with a full three dimensional model, which would allow for more realistic paleoclimate simulations.

The basic idea of variational thermodynamic principles such as the principle of \emph{Maximum Entropy Production}, which still remains to be proved, is that microscopic details are irrelevant, to a certain extent and depending on the question we are trying to answer, to macroscopic behaviour. The general process of not making any unjustified assumption on those microscopic details finds a nice formulation at equilibrium with the principle of \emph{Maximum Entropy} and the usual apparatus of equilibrium thermodynamics can be derived easily from the application of this principle to the classical ensembles of statistical mechanics \cite{Jaynes1957a}. But in spite of the conceptual similarity, the link between \emph{Maximum Entropy Production Principle} and \emph{Maximum Entropy Principle} still is not clear (see \cite{Dewar2003,Dewar2005} for an attempt to derive the former from the latter and \cite{Grinstein2007} for a comment). This study expresses the view that regardless of the veracity of the principle in full generality, it can certainly be useful in application in the field of climate sciences, as well as in many others.

\section*{\small Acknowledgements}
The authors would like to thank Jean-Louis Dufresne for fruitful discussions and advice, as well as Gilles Ramstein and Didier Roche for their stimulating remarks and constant encouragements.

\bibliographystyle{abbrv}
\bibliography{bibtexlib}

\clearpage

\begin{table*}
\begin{center}
\begin{tabular}{ccccc}
Profile & $u_{H_2O}$ ($g.cm^{-2}$) & T (K) & $t_a$ & $T_g^{MEP}$ (K)\\
\hline
Sub-Artic & 1.5 & 272 & 0.432 & 295.6 \\
Mid-Latitude & 2.3 & 283 & 0.436 & 295.9 \\
Tropical & 5.1 & 300. & 0.444 & 296.5 \\
\hline
\end{tabular}
\end{center}

\caption{Column-integrated water vapour density ($u_{H_2O}$), characteristic temperature (T) and resulting longwave atmospheric emissivity ($t_a$) for the three standard atmospheric profiles. The last column ($T_g^{MEP}$) gives the global mean temperature in the MEP state for pre-industrial surface albedo, with uniform atmospheric emissivity.}\label{tasensitivitytable}
\end{table*}

\begin{table*}
\begin{center}
\begin{tabular}{cccccc}
\hline
&Peixoto et al & PLASIM & HadCM3 & MEP Pre-industrial & MEP LGM\\
\hline
$\sigma_{SW}^{atm}$       & 258  & 255 &          &  216  &    217 \\
$\sigma_{SW}^{sfc}$         &  561 & 557&      \raisebox{1.5ex}{812}     &   667   &   650 \\
$\sigma_{LW}^{atm-sfc}$ & 24    & 7      & 11  &     7       &   7     \\
$\sigma_{LW}^{atm}$       &    -    &   28  &  39 &     -         &      -    \\
$\sigma_{M}$                     &  32 &  29  &  38  &     16    &     18  \\
$\sigma_{res}$                   &   17   &  7    &     12      &      0        &     0     \\
\hline
$\sigma_{total}$                    & 892  & 883 & 912 &     906   &    892  \\
\hline
\end{tabular}
\end{center}

\caption{Comparison of the entropy budget of the Earth computed from observations (\cite{Peixoto1991}, some values have been updated after \cite{kleidonlorenzbook}) and two models of different complexity (PLASIM: \cite{Fraedrich2008}, HadCM3: \cite{Pascale2009}), in $mW.m^{-2}.K^{-1}$. $\sigma_{res}$ designates all the other contributions to the total entropy production rate, including eventual imbalance.}\label{entropybudgettable}
\end{table*}

\clearpage

\end{document}